\documentclass[12pt]{article}
\usepackage[dvips]{graphicx}
\usepackage{textcomp}
\usepackage{amssymb}

\newcommand{\mc}{\multicolumn}

\begin{document}

\begin{titlepage}
\vskip0.5cm
\begin{flushright}
\end{flushright}

\vskip0.5cm
\begin{center}
{\Large\bf The thermodynamic Casimir effect in the neighbourhood of 
    the $\lambda$-transition: \\
\vskip0.2cm
 A Monte Carlo study of an improved three dimensional lattice model}
\end{center}

\centerline{
\large Martin Hasenbusch
}
\vskip 0.3cm
\centerline{\sl  Institut f\"ur Physik, Humboldt-Universit\"at zu Berlin}
\centerline{\sl Newtonstr. 15, 12489 Berlin, Germany}
\centerline{\sl
e--mail: \hskip 1cm
 Martin.Hasenbusch@physik.hu-berlin.de}
\vskip 0.4cm
\begin{abstract}
We study the thermodynamic Casimir effect in thin films in the 
three dimensional XY universality class.
To this end, we simulate  the improved two component
$\phi^4$ model on the simple cubic lattice. We use lattices
up to the thickness $L_0=33$. Based on the results 
of our Monte Carlo simulations we compute the universal finite 
size scaling function $\theta$ that characterizes the behaviour of the 
thermodynamic Casimir force in the neighbourhood of the critical point.
We confirm that leading corrections to the universal finite size 
scaling behaviour due to free boundary conditions can be expressed 
by an effective thickness $L_{0,eff} = L_0+ L_s$, with $L_s =1.02(7)$.
Our results are compared with experiments on films of $^4$He near 
the $\lambda$-transition, previous Monte Carlo simulations of the 
XY model on the simple cubic lattice and field-theoretic results.
Our result for the finite size scaling function $\theta$ is essentially
consistent with the experiments on films of $^4$He and the previous 
Monte Carlo simulations. 
\end{abstract}
{\bf Keywords:} $\lambda$-transition, Classical Monte Carlo
simulation, thin films, thermodynamic Casimir effect
\end{titlepage}

\section{Introduction}

In 1978  Fisher and de Gennes \cite{FiGe78} realized that there should
be a so called ``thermodynamic'' Casimir effect. This means that a force 
emerges when thermal fluctuations are restricted by a container. 
Thermal fluctuations extend to large scales in the neighbourhood of critical 
points. 
In the thermodynamic limit, in the neighbourhood of the critical point, various
quantities diverge following  power laws. E.g. the correlation length,
which measures the spatial extension of fluctuations, 
behaves as 
\begin{equation}
\label{xipower}
 \xi \simeq \xi_{0,\pm} t^{-\nu} 
\end{equation}
where $t = (T-T_c)/T_c$ is the reduced temperature and $T_c$ the critical
temperature. $\xi_{0,+} $ and 
$\xi_{0,-}$ are the amplitude of the correlation length in the high and 
low temperature phase, respectively. While $\xi_{0,+} $ and $\xi_{0,-}$ depend
on the microscopic details of the system, the critical exponent $\nu$ and 
the ratio $\xi_{0,+}/\xi_{0,-}$ are universal. This means that they assume
exactly the same values for all systems within a given universality class.
A universality class is characterized by the spatial dimension of the 
system, the range of the interaction and the symmetry of the order parameter.
The modern theory of critical phenomena is the Renormalization Group (RG).
For reviews see e.g. \cite{WiKo,Fisher74,Fisher98,PeVi02}. Here we consider
the XY universality class in three dimensions with short range interactions.
This universality class is of particular interest, since the 
$\lambda$-transition of $^4$He is supposed to share this universality class.
The most accurate experimental results for critical exponents and
universal amplitude ratios for a three dimensional system have been
obtained for this transition; for a review see \cite{BaHaLiDu07}. 

The critical behaviour is modified by
a confining geometry. If the system is finite in all directions,
thermodynamic functions have to be analytic. I.e. a singular
behaviour like eq.~(\ref{xipower}) is excluded.
As a remnant of such singularities there remains a peak in the
neighbourhood of the transition.  With increasing linear extension
the hight of the peak increases and the temperature of the maximum approaches
the critical temperature.
This  behaviour is described by the theory of finite size scaling (FSS).
For reviews see \cite{Barber,Privman}.  In general the physics in the 
neighbourhood of the transition is governed by the ratio $L_0/\xi$, where 
$L_0$ is the linear extension of the container and $\xi$ the correlation
length of the bulk system. Furthermore it depends on the geometry of the 
container and on the type of the boundary conditions that the container 
imposes on the order parameter. For a review on experimental studies of
$^4$He near the $\lambda$-transition in confining geometries 
see \cite{GaKiMoDi08}.

Here we study thin films. Thin films are finite in one direction
and infinite in the other two directions. In this case singular behaviour is
still possible. However the associated phase transition belongs to the
two-dimensional universality class. I.e. in the case of $U(1)$ symmetry,
a Kosterlitz-Thouless (KT) transition \cite{KT,Jo77,AmGoGr80} is expected.
In \cite{myKTfilm} we have confirmed the KT-nature of this transition
and have studied the scaling of the transition temperature with the
thickness of the film. Recently \cite{myheat} we determined 
the finite size scaling behaviour of the specific heat of thin films.
Here we investigate the thermodynamic Casimir force in thin films in the 
three dimensional XY universality class.

From a thermodynamic point of view,  the Casimir force per unit area is 
given by 
\begin{equation}
 F_{casimir} = - \frac{ \partial \tilde f_{ex} }{ \partial L_0} \;\;
\end{equation}
where $L_0$ is the thickness of the film and 
$ \tilde f_{ex}  =  \tilde f_{film} - L_0 \tilde f_{3D}$
is the excess free energy per area of the film, 
where $\tilde f_{film}$ is the free energy per area of the film and 
$\tilde f_{3D}$ the free energy density of the thermodynamic limit of the 
three dimensional system; see e.g. \cite{BrDaTo00}. 
Finite size scaling predicts that the Casimir force behaves as
\begin{equation}
\label{FSScasimir}
 F_{casimir} = \frac{k_B T}{ L_0^{3}} \theta(t [L_0/\xi_0]^{1/\nu})
\end{equation}
where $\theta(x)$ is a universal finite size scaling function.
\footnote{Following the literature, in eq.~(\ref{FSScasimir}) we shall use 
$\xi_{0,+}$ in the following.}
In \cite{GaCh99,GaScGaCh06} $^4$He films of thicknesses up to 588 $\AA$ 
have been studied. These experiments show clearly that the thermodynamic 
Casimir force is indeed present. Throughout it is negative. In the 
low temperature phase of the three dimensional bulk system it shows a
pronounced minimum. The data are essentially  consistent with 
the prediction eq.~(\ref{FSScasimir}). The minimum of $\theta(x)$
is located at $x=t (L_0/\xi_0)^{1/\nu} \approx - 5.5$. 

It has been a challenge for theorists to compute the finite size scaling
function $\theta(x)$. 
Krech and Dietrich \cite{KrDi92,KrDi92b} have computed it
in the high temperature phase using the
$\epsilon$-expansion up to O($\epsilon$). This result is indeed consistent 
with the measurements on $^4$He films. Deep in the low temperature phase,
the spin wave approximation should provide an exact result. It predicts a
negative non-vanishing value for $\theta(x)$.  However the experiments suggest 
a much  larger absolute value for $\theta(x)$ in this region.  Until recently
a reliable theoretical prediction for the minimum of $\theta(x)$ and its 
neighbourhood was missing.
Using a renormalized mean-field approach the authors of \cite{Kardar} 
have computed $\theta(x)$ for the whole temperature range. Qualitatively
they reproduce the features of the experimental result. However 
the position of the minimum is by almost a factor of 2 different from 
the experimental one. The value at the minimum is wrongly estimated by a 
factor of about 5.

Only quite recently Monte Carlo simulations of the
XY model on the simple cubic lattice \cite{VaGaMaDi07,Hu07,VaGaMaDi08}
provided results for $\theta(x)$ which essentially reproduce the experiments
on $^4$He films \cite{GaCh99,GaScGaCh06}. These simulations were 
performed  with lattices of a thickness up to $L_0=16$ \cite{Hu07} and 
up to $L_0=20$ \cite{VaGaMaDi08}. The authors of \cite{VaGaMaDi08} pointed 
out that for these lattice sizes corrections to scaling still play 
an important role. The purpose of the present work is to get accurate 
control over the leading corrections to scaling, allowing us to  compute
$\theta(x)$ with reliable error bars.

As the first step in this direction we simulate the improved two-component
$\phi^4$ model instead of the XY model. For a precise definition of these
models see the section below. This way we avoid leading corrections
to scaling which are $\propto L_0^{-\omega}$ with $\omega=0.785(20)$
\cite{recentXY}; similar values for the exponent are obtained with 
field-theoretic methods; for a review see e.g. \cite{PeVi02}.
In order to mimic the vanishing order parameter that is observed at
the boundaries of $^4$He films, Dirichlet boundary conditions with 
vanishing field are imposed. These lead to  corrections $\propto L_0^{-1}$ 
\cite{DiDiEi83}.  These corrections can be eliminated by
replacing the thickness $L_0$ by an effective one $L_{0,eff}=L_0+L_s$,
where $L_s=1.02(7)$ \cite{myKTfilm} for the model that we have simulated.
\footnote{
In the literature, replacing $L_0$ by $L_{0,eff} =L_0 + L_s$ to account for
surface corrections, was first discussed by Capehart and Fisher \cite{CaFi76}
in the context of the surface susceptibility of Ising films.}

This paper is organized as follows: First we define the model and the 
observables that we have measured. Next we discuss  the finite size 
scaling behaviour of the Casimir force. In particular, we discuss corrections 
to scaling  caused by the Dirichlet boundary conditions. We outline the 
method used to compute the Casimir force. We discuss the simulations
that have been performed and analyze our data. We compare our results
with experiments   \cite{GaCh99,GaScGaCh06}, previous Monte Carlo 
simulations \cite{Hu07,VaGaMaDi08} and the $\epsilon$-expansion 
\cite{KrDi92,KrDi92b}. Finally we summarize and conclude.

\section{The model and the observables}

We study the two component $\phi^4$ model on the simple cubic lattice.
We  label the sites of the lattice by
$x=(x_0,x_1,x_2)$. The components of $x$ might assume the values
$x_i \in \{1,2,\ldots,L_i\}$.  We simulate lattices
of the size $L_1=L_2=L$ and $L_0 \ll L$.  In 1 and 2-direction we employ
periodic boundary conditions and free boundary conditions in 0-direction.
This means that the sites with $x_0=1$ and $x_0=L_0$ have only five nearest
neighbours.
This type of boundary conditions could be interpreted as Dirichlet
boundary conditions with $0$ as value of the field at $x_0=0$ and $x_0=L_0+1$.
Note that viewed this way, the thickness of the film is $L_0+1$ rather
than $L_0$. This provides a natural explanation of the result $L_s=1.02(7)$
obtained in \cite{myKTfilm} and might be a good starting point for a
field theoretic calculation of $L_s$.
The Hamiltonian of the two component $\phi^4$ model, for a vanishing
external field, is given by
\begin{equation}
\label{hamiltonian}
{\cal H} = - \beta \sum_{<x,y>} \vec{\phi}_x \cdot \vec{\phi}_y
   + \sum_{x} \left[\vec{\phi}_x^2 + \lambda (\vec{\phi}_x^2 -1)^2   \right] 
\end{equation}
where the field variable $\vec{\phi}_x$ is a vector with two real components.
$<x,y>$ denotes a pair of nearest neighbour sites on the lattice.
The partition function is given by
\begin{equation}
Z =  \prod_x  \left[\int d \phi_x^{(1)} \,\int d \phi_x^{(2)} \right] \, \exp(-{\cal H}).
\end{equation}
Note that following the conventions of our previous work, e.g. \cite{ourXY},
we have absorbed the inverse temperature $\beta$ into the Hamiltonian.
\footnote{Therefore, following \cite{Fisher98} we actually should call it
reduced Hamiltonian.}
In the limit $\lambda \rightarrow \infty$ the field variables are fixed to
unit length; i.e. the XY model is recovered. For $\lambda=0$ we get the exactly
solvable Gaussian model.  For $0< \lambda \le \infty$ the model undergoes
a second order phase transition that belongs to the XY universality class.
Numerically, using Monte Carlo simulations and high-temperature series
expansions, it has been shown that there is a value $\lambda^* > 0$, where
leading corrections to scaling vanish.  Numerical estimates of $\lambda^*$
given in the literature are $\lambda^* = 2.10(6)$ \cite{HaTo99},
$\lambda^* = 2.07(5)$  \cite{ourXY} and most recently $\lambda^* = 2.15(5)$
\cite{recentXY}.  The inverse of the critical temperature $\beta_c$ has been
determined accurately for several values of $\lambda$ using finite size
scaling (FSS) \cite{recentXY}. We shall perform our simulations at
$\lambda =2.1$, since for this value of $\lambda$ comprehensive Monte
Carlo studies of the three-dimensional system in the low and the
high temperature phase have been performed
\cite{myKTfilm,recentXY,myAPAM,myamplitude}.
At $\lambda =2.1$ one gets $\beta_c=0.5091503(6)$ \cite{recentXY}.
Since  $\lambda =2.1$  is not exactly equal to $\lambda^*$, there are
still corrections $\propto L^{-\omega}$, although with a small amplitude.
In fact, following \cite{recentXY}, it should be by at least a factor
20 smaller than for the standard XY model.

In \cite{myKTfilm} we find for $\lambda=2.1$ by fitting the data for the
second moment correlation length in the high temperature phase
\begin{equation}
\label{xires}
 \xi_{2nd} = 0.26362(8) t^{-0.6717}   \times [1 +0.039(8) t^{0.527}
 - 0.72(4) t] \;\;,
\end{equation}
where $t=0.5091503-\beta$. Here we shall use $\nu=0.6717(1)$ \cite{recentXY}
as value of the critical exponent of the correlation length. Recent experiments
on the $\lambda$-transition of $^4$He suggest a slightly smaller value: 
$\nu=0.6709(1)$  \cite{lipa2003}. This discrepancy is however not crucial for
the present study.
Note that in the high temperature phase there is little difference between
$\xi_{2nd}$ and the exponential correlation length $\xi_{exp}$ which
is defined by the asymptotic decay of the two-point correlation function.
Following  \cite{ourXY}:
\begin{equation}
\lim_{t\rightarrow 0} \frac{\xi_{exp}}{\xi_{2nd}} = 1.000204(3)  \;\;,\;\;\;\; (t>0)
\;\;
\end{equation}
for the thermodynamic limit of the three-dimensional system.

\subsection{The internal energy and the free energy}
\label{defineE}
The reduced free energy density is defined as
\begin{equation}
\label{fdef1}
f = - \frac{1}{L_0 L_1 L_2} \log Z \;.
\end{equation}
I.e. compared with the free energy density $\tilde f$, a factor $k_B T$ is 
skipped.

Note that in eq.~(\ref{hamiltonian})  $\beta$ does not
multiply the second term. Therefore, strictly speaking, $\beta$ is not
the inverse of $k_B T$.
In order to study universal quantities it is not crucial how the transition
line in the $\beta$-$\lambda$ plane is crossed, as long as this path is
not tangent to the transition line.
Therefore, following computational convenience, we vary $\beta$ at fixed 
$\lambda$.
Correspondingly we define the (internal) energy density as the derivative of 
the reduced free energy density with respect to $\beta$.
Furthermore, to be consistent with our previous work \cite{myheat}, 
we multiply by $-1$:
\begin{equation}
\label{Edef1}
E = \frac{1}{L_0 L_1 L_2} \frac{\partial \log Z}{\partial \beta} \;.
 \end{equation}
 It follows
\begin{equation}
\label{Edef}
 E =  \frac{1}{L_0 L_1 L_2}
\left \langle  \sum_{<x,y>} \vec{\phi}_x \cdot \vec{\phi}_y \right \rangle \;,
\end{equation}
which can be easily determined in Monte Carlo simulations.  From
eqs.~(\ref{fdef1},\ref{Edef1}) it follows that the free energy density 
can be computed as
\begin{equation}
\label{integrateF}
 f(\beta) = f(\beta_0) - \int_{\beta_0}^{\beta} 
                       \mbox{d} \tilde \beta   E(\tilde \beta)   \;\;.
\end{equation}

\section{The finite size scaling behaviour of the thermal Casimir force}
Let us discuss the scaling behaviour of the reduced excess 
free energy.   
Since we study an improved model we ignore corrections
$\propto L_0^{-\omega}$ in the following.  We take into account leading
corrections due to the boundary conditions by replacing the thickness
$L_0$ of the film by $L_{0,eff}=L_{0}+L_s$ at the appropriate places. We 
split the free energies in singular (s) and non-singular (ns) parts:
\begin{eqnarray}
 f_{ex}(t,L_0) &=&  f_{film}(t,L_0) - L_0 f_{3D}(t) \nonumber  \\
 &=& f_{film,s}(t,L_0) + L_{0,eff,ns} f_{ns}(t) - L_{0} f_{3D,s}(t) 
 - L_0 f_{ns}(t)  \nonumber  \\
 &=& L_{0,eff}^{-2} h(x)
                          + L_s f_{3D,s}(t)
                          + L_{ns} f_{ns}(t) 
\end{eqnarray}
where $h(x)=L_{0,eff}^2 [f_{film,s}(t,L_0) - L_{0,eff} f_{3D,s}(t)]$ 
is a universal finite size scaling function and 
$x=t [L_{0,eff}/\xi_0]^{1/\nu}$. Following RG theory the non-singular part
is not affected by finite size effects. However it is not clear a priori 
how Dirichlet boundary conditions affect the non-singular part of the 
free energy. Therefore we allow for $L_{ns}=L_{0,eff,ns} - L_{0} \ne 0$ and
$L_{ns} \ne L_{s}$.
Taking the derivative with respect to $L_0$ we get the
thermodynamic Casimir force per area \cite{BrDaTo00}
\begin{equation}
\beta F_{casimir} = - \frac{\partial f_{ex}(t,L_0)}{\partial  L_0}=
 2  L_{0,eff}^{-3} h(x)
  - L_{0,eff}^{-3} \frac{1}{\nu} x h'(x)
  = L_{0,eff}^{-3} \theta(x) 
 \end{equation}
where  $\theta(x) = 2 h(x) - \frac{1}{\nu} x h'(x)$.  

\section{Computing the Casimir force on the lattice}
Here we follow essentially the approach of \cite{Hu07}. For an alternative
method see \cite{VaGaMaDi07,VaGaMaDi08}.
On the lattice the thickness $L_0$ assumes integer values. Therefore
we approximate the derivative for half-integer values of $L_0$ as
\begin{equation}
\label{derivue}
\left . \frac{\partial f_{ex}(\beta,L) }{\partial L} \right |_{L=L_0} 
\approx
\Delta f_{ex}(\beta,L_0) =f(\beta,L_0+1/2) - f(\beta,L_0-1/2) +  
f_{3D}(\beta) \;.
\end{equation}
Correspondingly we define
\begin{equation}
\label{deltaE}
\Delta E_{ex}(\beta,L_0) = E(\beta,L_0+1/2) - E(\beta,L_0-1/2) 
- E_{3D}(\beta)
\end{equation}
where $E(\beta,L_0)$ is the energy per area of a thin film
and $E_{3D}(\beta)$ the energy density of the three dimensional system.
Analogous to eq.~(\ref{integrateF}) we compute
\begin{equation}
\Delta f_{ex}(\beta) = - \int_{\beta_0}^{\beta}
\mbox{d} \tilde \beta \; \Delta E_{ex}(\tilde \beta) 
\end{equation}
where $\beta_0$ is chosen such that $\xi(\beta_0) \ll L_0$ and hence
the Casimir force vanishes.

\section{Numerical results}

In \cite{myheat} we have studied the specific heat of thin films in the 
two component $\phi^4$ model at $\lambda=2.1$. 
To this end, we had determined the energy density 
for the three dimensional thermodynamic limit and for films of the 
thicknesses $L_0=8$, $16$ and $32$ for a large number of $\beta$-values. 
In order to compute the derivative with respect to $L_0$, we have 
complemented these simulations by ones for $L_0=9$, $17$ and $33$. The 
Monte Carlo algorithm that has been used is the same as in \cite{myheat}:
An update-cycle is composed of a Metropolis sweep, a few overrelaxation
sweeps and single \cite{wolff} and wall \cite{HaPiVi99} cluster updates.  
One sweep means that a local update is performed at each site of the lattice
ones. As random number generator we have used the  SIMD-oriented Fast  
Mersenne Twister algorithm \cite{twister}.
In table \ref{statistics} we have summarized the statistics of our runs.
In total these simulations took about  3 years of CPU-time on a single core of a
Quad-Core Opteron(tm) 2378 CPU (2.4 GHz).

\begin{table}
\caption{\sl \label{statistics} We characterize our new simulations.
In the first column we give the thickness $L_0$ of
the film. In the second column we give the linear size $L=L_1=L_2$ of the 
lattice
in the other two directions. In the third and fourth column we give
the upper and lower bound of the interval in $\beta$
that has been simulated. In the fifth column we give the step size
$\Delta \beta$ that we used. E.g. $\beta_{min}=0.49$,
$\beta_{max}=0.519$ and $\Delta \beta=0.001$ means that
$\beta=0.49, 0.491, 0.492$,
$\ldots$, $0.519$ have been simulated. Finally, in the last column we give
the number of measurements (stat) that we have performed for each of the
simulations.
}
\begin{center}
\begin{tabular}{|c|r|l|l|l|c|}
\hline
 \mc{1}{|c|}{$L_0$} &
\mc{1}{|c|}{$L_1=L_2$} &
\mc{1}{|c|}{$\beta_{min}$}&
 \mc{1}{|c|}{$\beta_{max}$}&
  \mc{1}{|c|}{$\Delta \beta$} &
   \mc{1}{|c|}{stat} \\
\hline
 9  &   64  &   0.49         & 0.519  &  0.001      & $ 5 \times   10^5$\\
 9  &  128  &   0.52         & 0.527  & 0.001      & $ 2 \times   10^5$\\
 9  &  256  &   0.528        & 0.56   & 0.001      &  $ 10^5$ \\
 9  &  256  &   0.562        & 0.58   & 0.002      &  $ 10^5$ \\
 9  &  512  &   0.536        & 0.539  & 0.001      & $  10^5$ \\
 9  &  512  &   0.5395       & 0.548   & 0.0005     & $  10^5$ \\
 9  &  512  &   0.548        & 0.57   & 0.001       & $  10^5$ \\
 9  & 1024  &   0.539        & 0.548  & 0.0005     & $  10^5$ \\
\hline
17  &  256  &   0.527        & 0.55   &  0.001     & $ 2 \times   10^5$\\
17  &  512  &   0.5        & 0.512  &  0.001     & $  10^5$ \\
17  &  512  &  0.5125        & 0.529  &  0.0005    & $  10^5$ \\
17  &  512  &  0.53          & 0.55   &  0.001     & $  10^5$ \\
17  & 1024  &  0.5205        & 0.529  &  0.0005    & $ 8  \times 10^4$ \\
\hline
 33 &  256  &   0.502        & 0.50875& 0.00025    & $ 4 \times   10^5$\\
 33 &  512  &   0.509        & 0.5128&  0.0002     & $ 3 \times   10^5$ \\
 33 & 1024  &   0.513       &       &             & $    10^5$ \\
 33 & 1024  &   0.5132       &       &             & $ 8  \times 10^4$ \\
\hline
\end{tabular}
\end{center}
\end{table}

Using these data, we have computed $\Delta E_{ex}$ for $L_0=8.5$, $16.5$ 
and $32.5$.  One should note that the statistical error of 
$E(\beta,L_0+1/2) - E(\beta,L_0-1/2)$ is much larger than that of 
$E_{3D}(\beta)$.
In order to obtain $\Delta f_{ex}$ we have numerically integrated 
$\Delta E_{ex}$ using the trapezoidal rule:
\begin{equation}
\label{integration}
-\Delta f_{ex}(\beta_n) \approx \sum_{i=0}^{n-1} \frac{1}{2} (\beta_{i+1}-\beta_i)
   \left(\Delta E_{ex}(\beta_{i+1}) + \Delta E_{ex}(\beta_{i}) \right) 
\end{equation}
where $\beta_{i}$ are the values of $\beta$ we have simulated at. They
are ordered such that $\beta_{i+1} > \beta_i$ for all $i$.  
We have chosen $\beta_0=0.49$, $0.5$ and $0.505$ for 
$L_0=8.5$, $16.5$ and $32.5$.  We find that 
$\Delta E_{ex}$ is equal to zero within error bars up to values of $\beta$ 
that are slightly larger than the $\beta_0$ that we have chosen.

The estimate obtained from the integration is affected by statistical 
and systematical errors. The statistical one can be easily computed, since 
the $\Delta E_{ex}$ are obtained from independent simulations:
\begin{eqnarray}
 \epsilon^2 (-\Delta f_{ex}(\beta_n)) 
 &=& \frac{(\beta_{1} - \beta_{0})^2}{4} \epsilon^2 [\Delta E_{ex}(\beta_0)]
  + \frac{(\beta_{n} - \beta_{n-1})^2}{4} \epsilon^2 [\Delta E_{ex}(\beta_n)]
  \nonumber \\
 &+& \sum_{i=1}{n-1} \frac{(\beta_{i+1} - \beta_{i-1})^2}{4} 
      \epsilon^2 [\Delta E_{ex}(\beta_i)]
\end{eqnarray}
where $\epsilon^2$ denotes the square of the statistical error.

In order to estimate the error due to the finite step size
$\beta_{i+1}-\beta_{i}$ we have redone the integration, skipping every
second value of $\beta$; i.e. doubling the step size. In all three cases
(i.e. $L_0=8.5$, $16.5$ and $32.5$),  the results were consistent
within the statistical errors. Therefore we are confident that the systematical
error due to the finite step size is smaller than the statistical one.

\begin{figure}
\begin{center}
\scalebox{0.62}
{
\includegraphics{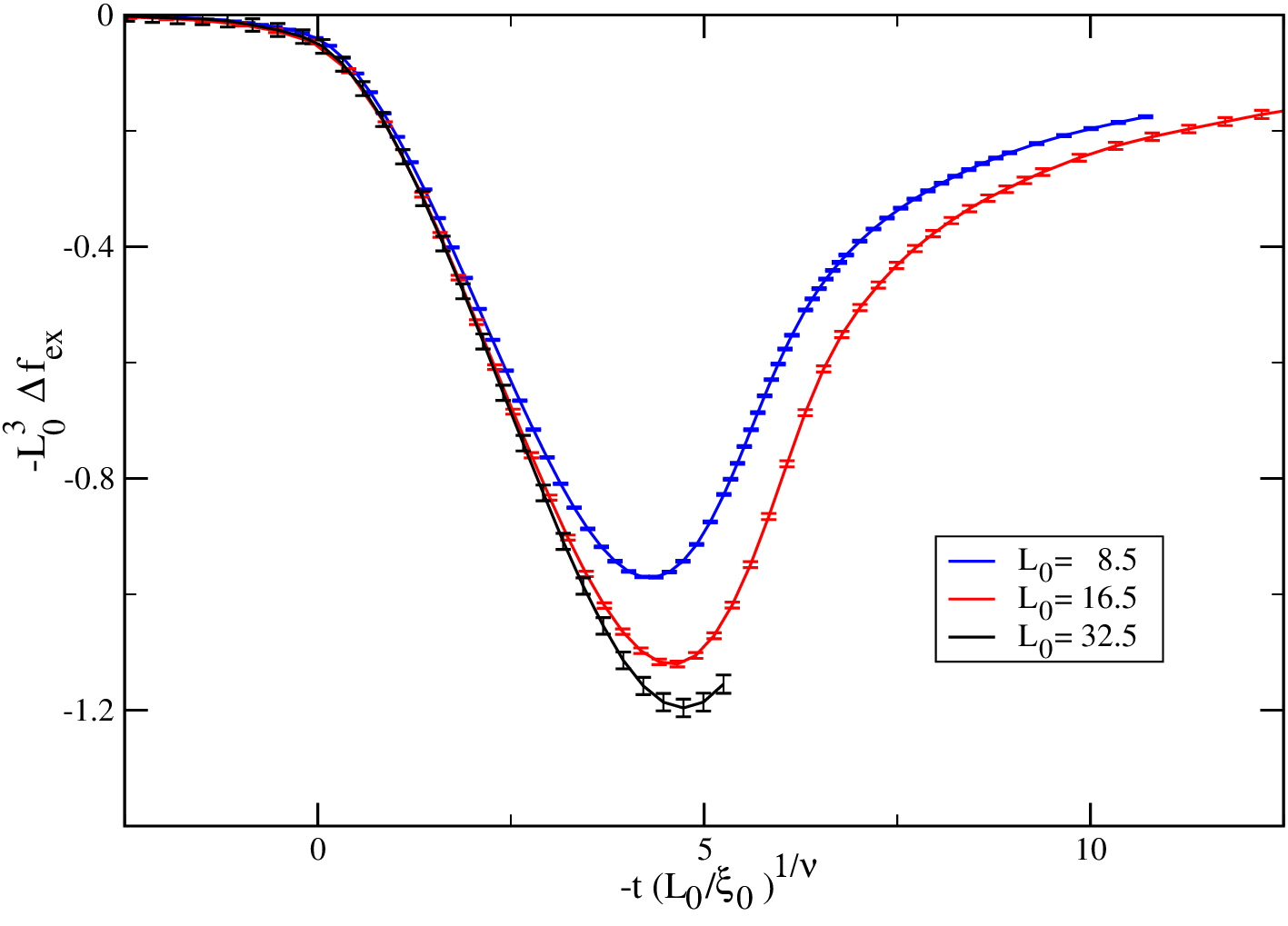}
}
\end{center}
\caption{
\label{nocorr}  We plot
$- L_{0}^3 \Delta f_{ex}$
as a function of $- t (L_0/\xi_0)^{1/\nu}$
for $L_0=8.5$,  $16.5$ and $32.5$, where we use $\nu=0.6717$ and 
$\xi_0 =0.26362$.
For a discussion see the text.
}
\end{figure}

In figure \ref{nocorr} we have plotted $- L_0^3 \Delta f_{ex}$ as a function
of $- t [L_0/\xi_0]^{1/\nu}$, where we have used $\nu=0.6717$ and 
$\xi_0=0.26362$, eq.~(\ref{xires}). We find that throughout the function  
assumes a negative value. In all cases it has a single minimum at 
$t [L_0/\xi_0]^{1/\nu} \approx - 5$.
The position of the minimum  $\beta_{min}(L_0)$ can be easily determined: 
It is given by the 
zero of $\Delta E_{ex}$.  We have computed $\beta_{min}(L_0)$
by linearly fitting $\Delta E_{ex}$ in the neighbourhood of the minimum.
In addition to $L_0=8.5$, $16.5$ and $32.5$ we performed simulations for
$L_0=6.5$, $7.5$, $9.5$, $12.5$ and $24.5$ at a few values of $\beta$ 
in the neighbourhood of $\beta_{min}$.  
Our results are summarized in table \ref{minimum}.

\begin{figure}
\begin{center}
\scalebox{0.62}
{
\includegraphics{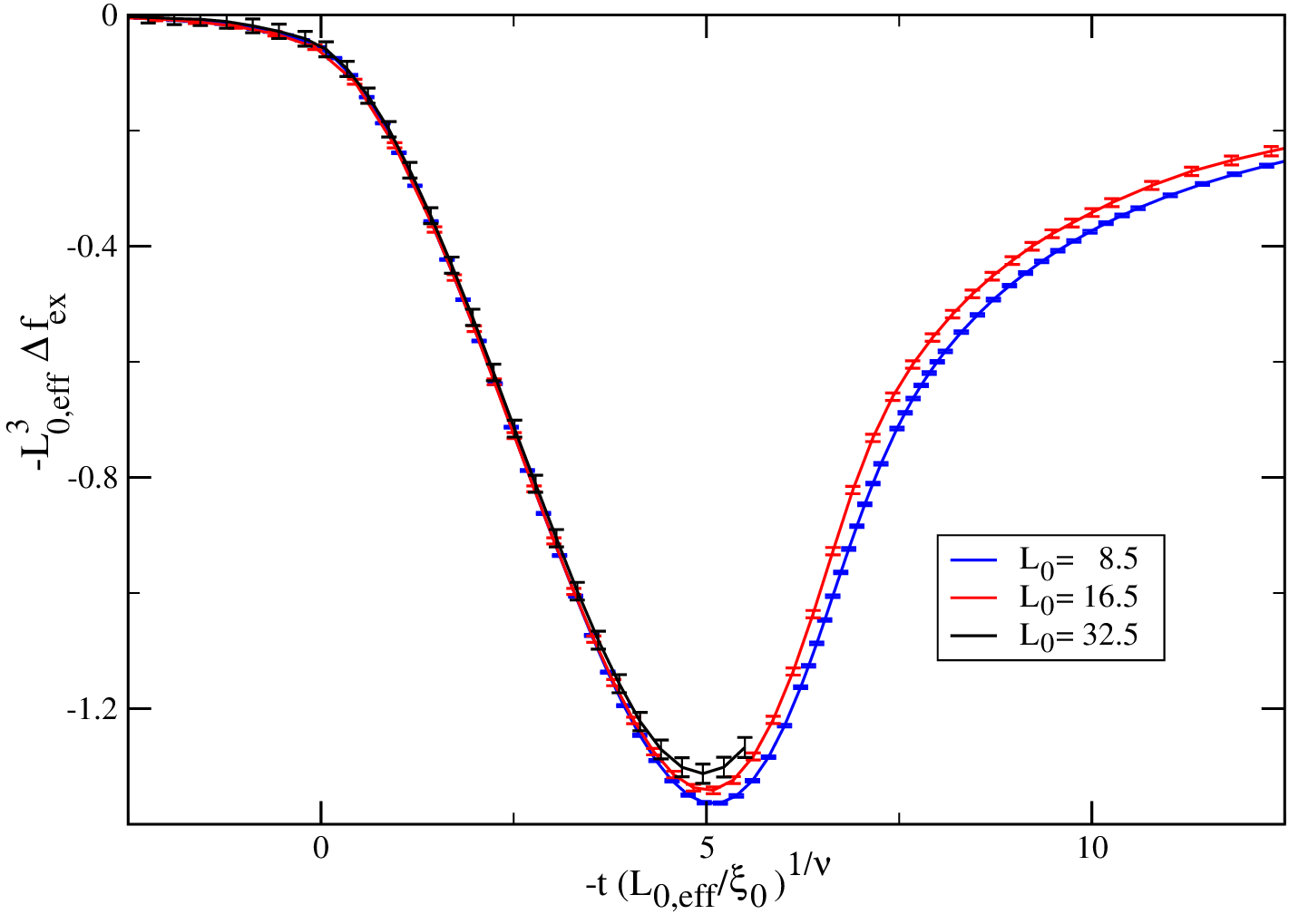}
}
\end{center}
\caption{
\label{leff}  We plot
$- L_{0,eff}^3 \Delta f_{ex} $
as a function of $-t (L_{0,eff}/\xi_0)^{1/\nu}$
for $L_0=8.5$,  $16.5$ and $32.5$, where we use $\nu=0.6717$,
$\xi_0 =0.26362$ and $L_{0,eff}=L_0+L_s$ with $L_s=1.02$.
For a discussion see the text.
}
\end{figure}

The curves for $L_0=8.5$, $16.5$ and $32.5$ plotted in 
figure \ref{nocorr} do not fall on top of each other. E.g. both the 
position and the value of the minimum are quite different for different 
$L_0$. 
In order to take corrections into account we have replaced $L_0$ by 
$L_{0,eff}= L_0 + L_s$, where $L_s = 1.02(7)$ \cite{myKTfilm}.  To this end, 
in figure \ref{leff} we have plotted $- L_{0,eff}^3 \Delta f_{ex}$ as a function
of $-t [L_{0,eff}/\xi_0]^{1/\nu}$, where we have used the central value of 
the shift $L_s=1.02$. Now indeed the distance between the curves for different
$L_0$ is much reduced compared with figure \ref{nocorr}. The results
for $L_{0}=16.5$ and $L_{0}=32.5$ are almost consistent within error bars.
Note that using $L_s =0.95$ the matching of the data for different 
$L_0$ seems to be better than for $L_s=1.02$.

Let us discuss in more detail the results obtained for the minimum of
the finite size scaling function $\theta(x)$. Using the numbers given 
in the third column of table \ref{minimum} and $L_s=1.02$ we get  
$-\Delta f_{ex,min} L_{0,eff}^3 =$  $-1.365(3)$, $-1.341(6)$ and $-1.311(19)$
for $L_0=8.5$, $16.5$ and $32.5$, respectively. Using instead $L_s=0.95$ we
get $-1.335(3)$, $-1.325(6)$ and $-1.302(19)$ for $L_0=8.5$, $16.5$ and $32.5$, 
respectively. As our final result we take the one obtained from 
$L_0=32.5$ and $L_s=1.02$:  
\begin{equation}
\theta_{min} = -1.31(3) \;\;, 
\end{equation}
where the error that is quoted
takes into account the statistical error and the uncertainty of $L_s$.

\begin{table}
\caption{\sl \label{minimum} We give the position $\beta_{min}$ of the 
minimum of the Casimir force and its value $-\Delta f_{ex,min}$ as a 
function of the thickness $L_0$.
}
\begin{center}
\begin{tabular}{|r|c|l|}
\hline
 \mc{1}{|c}{$L_0$} &
 \mc{1}{|c}{$\beta_{min}$} &
 \mc{1}{|c|}{$- \Delta f_{ex,min}$} \\
\hline
   6.5 & 0.54432(2) &  \\
   7.5 & 0.53814(2) &  \\
   8.5 & 0.53354(2) &--0.001582(3) \\
   9.5 & 0.53010(2) &  \\
  12.5 & 0.52348(2) &  \\
  16.5 & 0.51886(2) &--0.0002494(11) \\
  24.5 & 0.51463(2) & \\
  32.5 & 0.51279(2) & --0.0000348(5) \\     
\hline
\end{tabular}
\end{center}
\end{table}

Next we have fitted our results for $\beta_{min}$ with the ansatz
\begin{equation}
\label{fitmin}
 t_{min} (1 + c t_{min})  (L_{0,eff}/\xi_0)^{1/\nu}  = x_{min}
\end{equation}
where $t_{min} = \beta_c- \beta_{min}$. We have used 
$\nu=0.6717$, $\xi_0=0.26362$, $\beta_c=0.5091503$ and $L_s=1.02$ as input
and $c$ and $x_{min}$ as parameters of the fit. The term $(1 + c t_{min})$
parametrizes analytic corrections.  We did not include a correction
with the exponent $\theta' \approx 1.2$ \cite{RG}, since within 
the accuracy of our data they can not be discriminated from the leading 
analytic correction.
The results of these fits are summarized in table \ref{fitminimum}.
The $\chi^2$/d.o.f. is reasonably small starting from $L_{0,min}=6.5$,
where all thicknesses $L_0$ with $L_0 \ge L_{0,min}$ are included 
into the fit.
Also the estimates for $x_{min}$ and $c$ do not change much as 
$L_{0,min}$ is changed. 
In order to check the dependence of the results on the value of $L_s$ 
we have repeated the fit for  $L_{0,min}=8.5$ using $L_s=0.95$ instead of
\begin{table}
\caption{\sl \label{fitminimum}  We fit our results for $\beta_{min}$ with 
ansatz~(\ref{fitmin}). $L_{0,min}$ is the smallest thickness of the film
that is included into the fit. 
}

\begin{center}
\begin{tabular}{|c|l|l|c|}
\hline
 \mc{1}{|c}{$L_{0,min}$} &
  \mc{1}{|c}{$x_{min}$} &
   \mc{1}{|c}{$c$} &
   \mc{1}{|c|}{$\chi^2/$d.o.f.} \\
\hline
    6.5       &  --4.942(6)   & 1.20(4) & 1.47   \\
    7.5       &  --4.945(8)   & 1.17(7)  & 1.72  \\
    8.5       &  --4.956(10)  & 1.04(10) & 1.32 \\
    9.5       &  --4.952(12)  & 1.10(12) & 1.64 \\
\hline
\end{tabular}
\end{center}
\end{table}
$L_0=1.02$.  We get  
the $x_{min}=-4.943(10)$, $c=0.70(10)$ with $\chi^2$/d.o.f.$=1.12$.
As our final result we take
\begin{equation}
 x_{min} = - 4.95(3)
\end{equation}
where the error bar covers the statistical error and the uncertainty of
$L_s$.

Our result for $x_{min}$ can be compared with those given in the 
literature. The experimental works \cite{GaCh99} give 
$x_{min} = - 5.45(12)$ (no final result for $\theta_{min}$ is quoted) and
\cite{GaScGaCh06} $x_{min} = - 5.7(5)$ and $\theta_{min}= -1.30(3)$.
In \cite{GaCh99} and \cite{GaScGaCh06} the convention $x= t L_0^{1/\nu}$ is 
used. In order to convert to $x= t [L_0/\xi_0]^{1/\nu}$ we have used 
$\xi_0=1.422 \AA$ for
$^4$He at vapour pressure as discussed in section 4.2 of \cite{myheat}.

In his Monte Carlo study of the XY model on the simple cubic lattice,  
Hucht \cite{Hu07} finds $x=-5.3(1)$ and $\theta_{min}= -1.35(3)$.
The authors of \cite{VaGaMaDi08} have used two different ans\"atze
for the corrections to scaling. Using the first one, they arrive at
$x_{min} = -5.43(2)$ and $\theta_{min}= -1.396(6)$ and using the second 
one at $x_{min} = -5.43(2)$ and $\theta_{min}= -1.260(5)$.
Note that the ans\"atze used by \cite{VaGaMaDi08}, see eqs.~(20,21,22,23) of
\cite{VaGaMaDi08},
provide only an overall $x$-independent factor; therefore  
they do not  allow for any correction to scaling of $x_{min}$.

Our results for $\theta_{min}$ is in good agreement with both the experiment
\cite{GaScGaCh06} as well as with previous Monte Carlo studies of the 
XY model \cite{Hu07,VaGaMaDi08}. On the other hand, the position of the
minimum $x_{min}$ differs by several times the quoted error bar from both the 
experiment \cite{GaCh99,GaScGaCh06} as well as from Monte Carlo studies of the 
XY model \cite{Hu07,VaGaMaDi08}.

\begin{figure}
\begin{center}
\scalebox{0.62}
{
\includegraphics{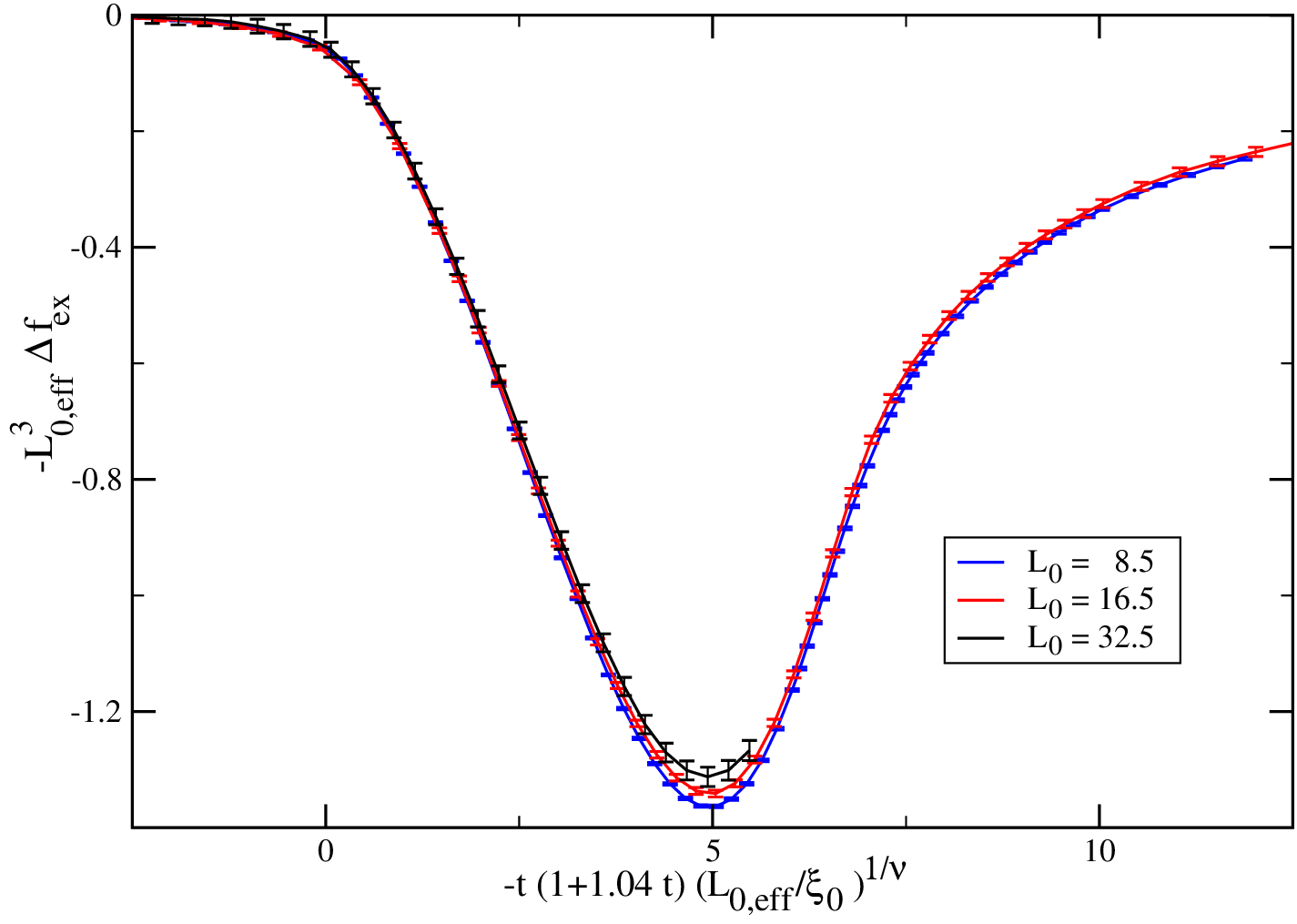}
}
\end{center}
\caption{
\label{analplot}  We plot
$- L_{0,eff}^3 \Delta f_{ex} $ is plotted
as a function of $-t (1+1.04 t) (L_{0,eff}/\xi_0)^{1/\nu}$
for $L_0=8.5$,  $16.5$ and $32.5$, where we use $\nu=0.6717$,
$\xi_0 =0.26362$ and $L_{0,eff}=L_0+L_s$ with $L_s=1.02$.
For a discussion see the text.
}
\end{figure}

Finally, in figure \ref{analplot}  we take into account the analytic 
corrections that we have detected fitting the position $t_{min}$ of the 
minimum of the Casimir force.  To this end, we have replaced the 
argument $t (L_{0,eff}/\xi_0)^{1/\nu}$ by 
$t (1+1.04 t) (L_{0,eff}/\xi_0)^{1/\nu}$.
The coefficient of the analytic correction is taken from the fit where 
we have fixed $L_s=1.02$ and $L_{0,min}=8.5$.  Now we find an almost 
perfect match between the curves obtained from $L_0=8.5$, $16.5$ and $32.5$.

\subsection{Comparison with other theoretical approaches}
\begin{figure}
\begin{center}
\scalebox{0.62}
{
\includegraphics{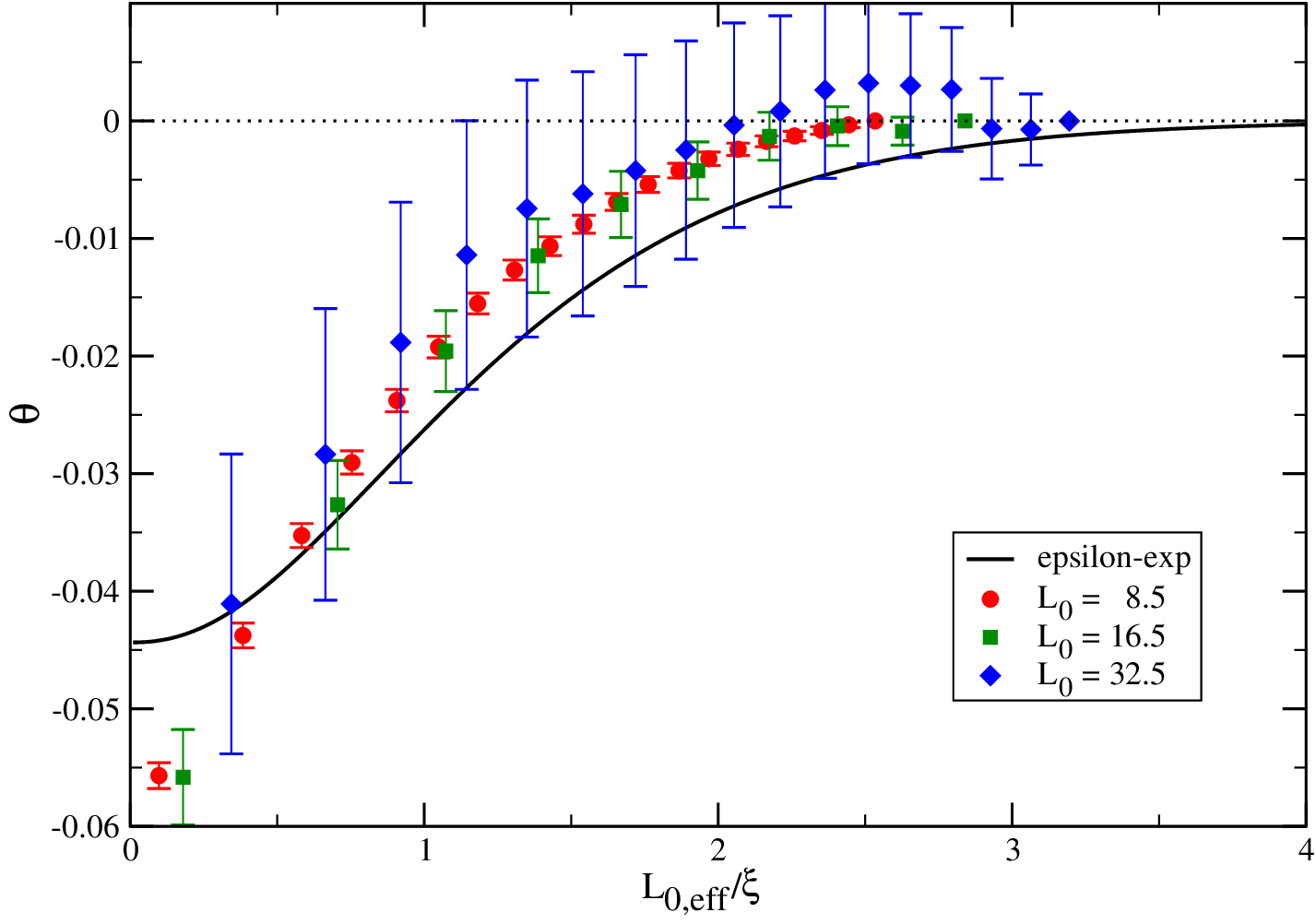}
}
\end{center}
\caption{
\label{epsilon_plot}  We plot the finite size scaling function 
$\theta$ as a function of $L_0/\xi$ in the high 
temperature phase obtained by Krech and Dietrich 
\cite{KrDi92,KrDi92b} using the $\epsilon$-expansion.
For comparison  we give our results obtained for $L_0=8.5$, $L_0=16.5$
and $L_0=32.5$. In the case of our data the leading boundary correction
is taken into account by replacing $L_0$ by $L_{0,eff}=L_0+L_s$ with 
$L_s=1.02$. For a discussion see the text.
}
\end{figure}

Krech and Dietrich \cite{KrDi92,KrDi92b} have computed the finite size 
scaling function $\theta$ in the high temperature phase using the 
$\epsilon$-expansion up to O($\epsilon$).  In figure \ref{epsilon_plot} we plot 
their result for the XY universality class  ($N=2$) setting $\epsilon=1$.
For comparison we plot our results for $L_0=8.5$, $L_0=16.5$ and $L_0=32.5$.
We have taken into account leading boundary corrections by 
replacing $L_0$ by $L_{0,eff} = L_0 + L_s$, where we have taken $L_s=1.02$.

Comparing with the $\epsilon$-expansion  we can estimate the systematical
error caused by setting $\Delta f_{ex}(\beta_0)=0$ in eq.~(\ref{integration}):
We read off from the $\epsilon$-expansion  that
$\theta\approx -0.0035, -0.0022, -0.0015$ for our choices of $\beta_0$ for
$L_0=8.5$, $16.5$ and $32.5$. Taking into account this
error, we see a good agreement
of our Monte Carlo results and the $\epsilon$-expansion down to 
$L_0/\xi \approx 1$. The curve obtained from
the $\epsilon$-expansion flattens as 
the critical point is approached. At the critical point the slope vanishes.
In contrast, in our case, the curve steepens as the critical point is 
approached.

The authors of \cite{Kardar} have computed the finite size scaling function
$\theta$ using a renormalized  mean field approach.  While this approach 
correctly reproduces  qualitative features of the finite size 
scaling function $\theta$ it fails to give quantitatively accurate 
results.
In particular the authors of  \cite{Kardar}  find 
$x_{min}= - \pi^2 \approx - 9.8696$ and $\theta_{min} \approx -6.92$. 
I.e. The position of the minimum is overestimated by about a factor of 
2 and its value by a factor of about 5. 

The authors of \cite{BiBhSaBhHu08} incorporate fluctuation effects into
their mean-field analysis. Qualitatively they get the finite size
scaling function $\theta$ for the whole range of $x$ right.
They adjust the two parameters  of their solution for the low temperature 
phase (eq.~(17)  of \cite{BiBhSaBhHu08} ) such that the minimum of 
$\theta$ found in the experiments \cite{GaCh99,GaScGaCh06} is reproduced.
We did the same exercise, adjusting to our result for the minimum of 
$\theta$. We find no good match between  $\theta$ computed in 
\cite{BiBhSaBhHu08} and ours. Our minimum is much
more peaked than that of \cite{BiBhSaBhHu08}.

\subsection{Comparison with experimental results}
\begin{figure}
\begin{center}
\scalebox{0.62}
{
\includegraphics{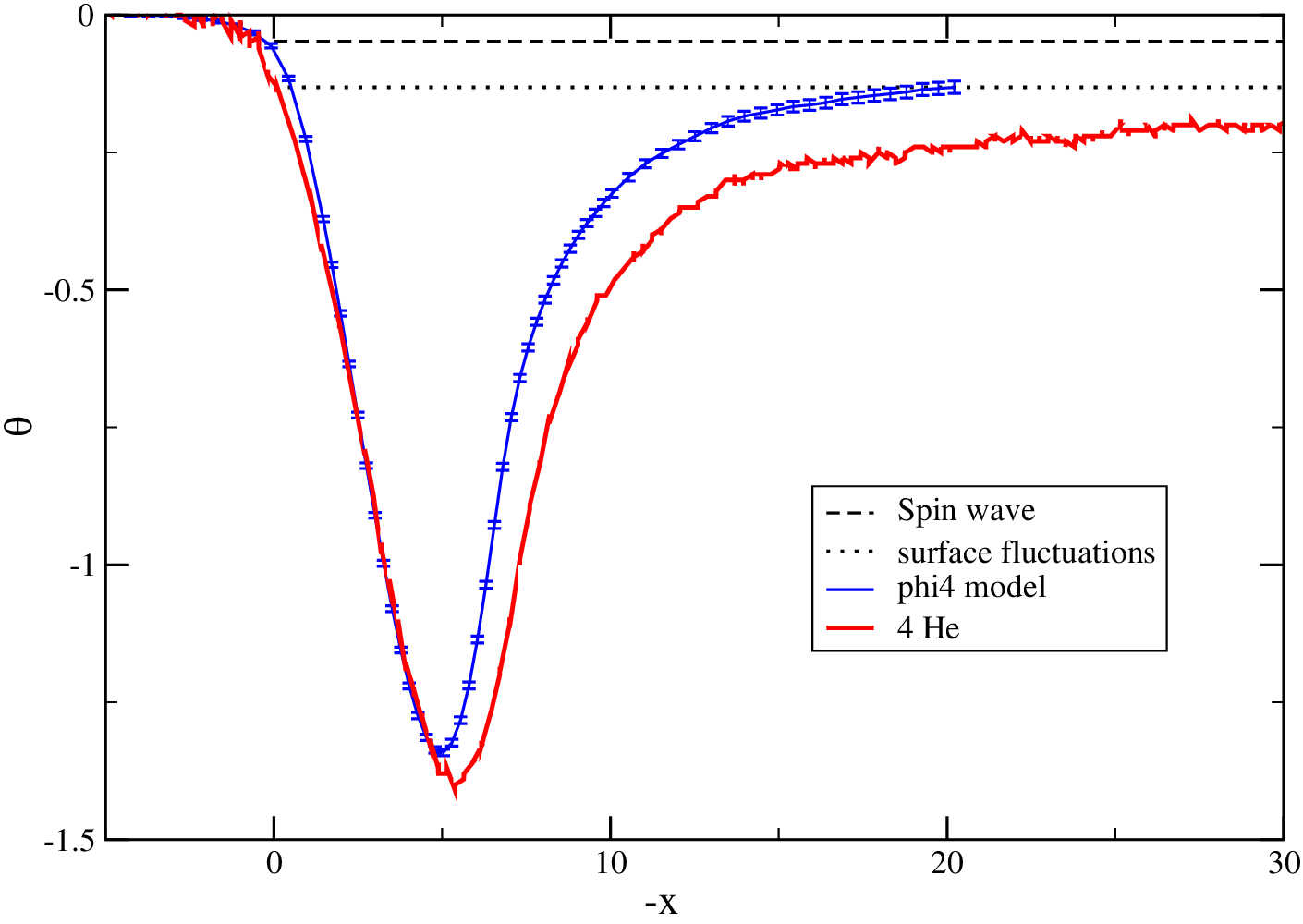}
}
\end{center}
\caption{
\label{experimentplot}  We plot the finite size scaling function
$\theta(x)$ obtained from an experiment on a thin film of $^4$He
\cite{GaCh99,webpage}, where $x=(T/T_{\lambda}-1) (L_0/\xi_0)^{1/\nu}$ 
with $\xi_0=1.422 \AA$. For comparison we give our result obtained 
from $L_0=16.5$, where we have taken into account boundary and analytic 
corrections as in figure \ref{analplot}. Furthermore we give the 
asymptotic value $-0.04783$ for $x \rightarrow -\infty$ (dashed line). 
The authors of \cite{Kardar} have argued that fluctuations of the surface
of the $^4$He film gives an additional contribution  to the Casimir force.
The corresponding asymptotic value $-0.1315$ is given by the dotted line.
For a discussion see the text.
}
\end{figure}

Finally we compare our result for the finite size scaling function 
$\theta$ with experiments \cite{GaCh99,GaScGaCh06}. 
In \cite{GaCh99} films of thicknesses  between $298$ and $588 \AA$ have
been studied. In figure 
\ref{experimentplot} we have plotted the data  obtained from capacitor 1
which corresponds to the thickness $575 \AA$ of the film at temperatures
$T>T_{\lambda}$.  This set of data is the smoothest among the five sets
given in \cite{GaCh99,webpage}. The results of 
\cite{GaScGaCh06} are, in the range of temperatures we are interested in,
less precise than those of \cite{GaCh99}.  In the tables provided by the 
authors \cite{webpage} the 
finite size scaling function $\theta$ is given as a function of 
$(T/T_{\lambda}-1) L_0^{1/\nu}$. In order to compare with our results
we have converted this to $(T/T_{\lambda}-1) (L_0/\xi_0)^{1/\nu}$, using
$\xi_0=1.422 \AA$.
For comparison we give our result obtained 
from $L_0=16.5$, where we have taken into account the boundary correction 
by replacing $L_0$ by $L_{0,eff}$ and the leading analytic correction as 
discussed above.  Furthermore, we give the asymptotic value  
\cite{LiKa91,Kardar0}
\begin{equation}
\label{spinwave}
\lim_{x \rightarrow \infty}  \theta(x)  = - \frac{\zeta(3) }{8 \pi} \approx
- 0.04783
\end{equation}
obtained from the spin wave  approximation. 

As already observed  by the authors of \cite{Hu07,VaGaMaDi08} there is 
a qualitative agreement among the result obtained from Monte 
Carlo simulations of lattice models and the experiment.  There is a
reasonable agreement of the position of the minimum $x_{min}$, as already 
discussed above.  For $x<x_{min}$  our result is in good agreement with that
of the experiment. In contrast for $x>x_{min}$  the experimental value of
$\theta$ is clearly smaller than ours.  
For $x \approx - 30$  the experimental result (capacitor 1 of \cite{GaCh99}) 
assumes $\approx -0.19$ and is decreasing again for smaller values of 
$x$.
This is clearly different from the prediction~(\ref{spinwave}) of the spin
wave approximation.

In ref. \cite{Kardar0} it was argued that this discrepancy could be 
explained by fluctuations of the surface resulting in  
\begin{equation}
\label{surfacefluct}
\lim_{x \rightarrow - \infty} \tilde \theta(x)  = - \frac{11 \zeta(3) }{32 \pi} 
\approx - 0.1315  \;.
\end{equation}
This goes indeed in the right direction, can however not fully explain 
the difference between the experimental result and the theoretical 
prediction~(\ref{spinwave}).


\section{Summary and Conclusions}
We have simulated the improved
two component $\phi^4$ model on the simple cubic lattice. This model shares
with the $\lambda$-transition of $^4$He the three
dimensional XY universality class. We consider the thin film 
geometry. In order to mimic the vanishing order parameter at the 
surface of $^4$He films near the $\lambda$-transition, we impose Dirichlet
boundary conditions with a vanishing field. 

Restricting the system to a finite geometry leads to an effective force 
called thermodynamic Casimir force. Its behaviour in the neighbourhood 
of the critical point is characterized by a universal finite size 
scaling function. We have computed this function and have compared 
our result with that obtained from experiments on films of $^4$He 
\cite{GaCh99,GaScGaCh06,webpage}, previous Monte Carlo simulations 
of the XY model on the simple cubic lattice \cite{VaGaMaDi07,Hu07,VaGaMaDi08},
field theoretic methods \cite{KrDi92,KrDi92b} and mean field
approaches \cite{Kardar,BiBhSaBhHu08}. 

The thermodynamic Casimir force is given as minus the derivative 
of the excess free energy of the film with respect to its thickness $L_0$.
On the lattice, this is approximated by the finite difference of films
of the thickness $L_0 +1/2$ and $L_0 -1/2$, where $L_0 +1/2$ is integer.

In general it is impossible to compute free energies from a single 
Monte Carlo simulation. To circumvent this problem, divide and conquer 
strategies are employed. Different strategies have been proposed in 
\cite{VaGaMaDi07,VaGaMaDi08} and \cite{Hu07}.  Here we 
essentially follow \cite{Hu07}: We compute the derivative of the 
excess energy with respect to $L_0$
for a dense grid of temperature values in the neighbourhood of the critical
point.  The corresponding result for the free energy is then obtained 
by numerical integration. 

As discussed in ref. \cite{VaGaMaDi08}, corrections to scaling 
are numerically quite large for the thicknesses that can be studied 
at present. The main purpose of the present  work is to get better 
control over these corrections as in previous work \cite{Hu07,VaGaMaDi08}.

To this end, we have studied the $\phi^4$ model at $\lambda=2.1$. In this
model leading corrections to scaling (and finite size scaling) which 
are $\propto L_0^{-\omega}$, with $\omega=0.785(20)$, are suppressed 
at least by a factor of 20 compared  with the XY model \cite{recentXY}.

Boundary effects lead to corrections $\propto L_0^{-1}$. These 
corrections can be cast into the form $L_{0,eff} = L_0 + L_s$. 
In \cite{myKTfilm} we have determined 
$L_s=1.02(7)$ for the two component $\phi^4$ model at $\lambda=2.1$ by
using a finite size 
scaling study at the critical point of the three dimensional system.
We have verified that this choice of $L_s$ indeed 
eliminates the leading boundary correction in  the scaling 
of the temperature of the Kosterlitz-Thouless transition \cite{myKTfilm}
and the specific heat of thin films \cite{myheat}.  Also here we 
confirm that corrections can be essentially eliminated by 
replacing $L_0 $ by $L_{0,eff} = L_0 + L_s$ with $L_s=1.02(7)$. 
Remaining discrepancies can be fitted by analytic corrections. 

Essentially we confirm the results obtained by previous Monte 
Carlo simulations of the three dimensional XY model \cite{Hu07,VaGaMaDi08} 
for the finite size scaling function $\theta$. 
The main discrepancy with these previous works is the 
position of the minimum $x_{min}=- 4.95(3)$  compared with $x_{min}=-5.3(1)$  
\cite{Hu07} and $x_{min}=-5.43(2)$ \cite{VaGaMaDi08}.

We should note that the Monte Carlo simulation of lattice models
is at the moment  the only theoretical method that allows for 
a quantitatively accurate calculation of $\theta$
in the low temperature phase not too far from the critical point.
The $\epsilon$-expansion gives correctly the behaviour in the high 
temperature phase. The spin wave approximation gives the exact result
in the limit $x \rightarrow -\infty$.
The mean field calculation of \cite{Kardar} reproduces only qualitatively 
the features of the scaling function. Quantitatively it is not 
satisfactory: the position of the minimum is wrongly estimated by a 
factor of almost $2$ and its value by a factor of about 5.

Qualitatively the Monte Carlo studies of lattice models 
nicely reproduce the finite 
size scaling function obtain from the experimental data 
\cite{GaCh99,GaScGaCh06,webpage} for films
of $^4$He.  For $x > x_{min}$ there is a very good quantitative agreement 
between the two.  In contrast, for $x < x_{min}$ the value obtained from 
the experiment is clearly smaller than that of the Monte Carlo studies. 
At large values of $-x$ the spin wave approximation should become exact.
Also in this regime, the experiments produce numbers that are too small
compared with the theoretical one.  The authors of \cite{Kardar}  
explain this  discrepancy by fluctuations of the surface of the 
$^4$He film.  Their result indeed reduces but not completely eliminates
the difference between experiment and theory. Therefore
it might be interesting to perform Monte Carlo simulations
of a lattice model that  incorporates fluctuations of the surface of the 
film.

\section{Acknowledgements}
This work was supported by the DFG under the grant No HA 3150/2-1.


\begin{thebibliography}{99}
\bibitem{FiGe78}
Fisher M E and de Gennes P-G,
{\sl Phenomena at the walls in a critical binary mixture}, 1978 
CR\ Acad.\ Sci.\ Paris {\bf B 287}  207

\bibitem{WiKo}
Wilson K G and Kogut J,
{\sl The renormalization group and the $\epsilon$-expansion}, 1974
Phys.\ Rep.\ C {\bf 12} 75

\bibitem{Fisher74}
Fisher M E, {\sl The renormalization group in the theory of critical behavior},
1974  Rev.\ Mod.\ Phys.\ {\bf 46} 597

\bibitem{Fisher98}
Fisher M E,
{\sl Renormalization group theory:
Its basis and formulation in statistical physics}, 1998
 Rev.\ Mod.\ Phys.\ {\bf 70} 653

\bibitem{PeVi02}
Pelissetto A and Vicari E,
{\sl Critical Phenomena and Renormalization-Group Theory}, 2002
Phys.\ Rept.\ {\bf 368} 549 [arXiv:cond-mat/0012164]

\bibitem{BaHaLiDu07}
Barmatz M, Hahn I, Lipa J A, and Duncan R V,
{\sl Critical phenomena in microgravity: Past, present, and future}, 2007
Rev.\ Mod.\ Phys.\ {\bf 79} 1

\bibitem{Barber}
M. N. Barber {Finite-size Scaling} in {\sl Phase Transitions and Critical Phenomena, Vol. 8,}
eds. C. Domb and J. L. Lebowitz, (Academic Press, 1983)

\bibitem{Privman}
{\sl Finite Size Scaling and Numerical Simulation of Statistical Systems,}
ed. V. Privman,
(World Scientific, 1990).

\bibitem{GaKiMoDi08}
Gasparini F M, Kimball M O, Mooney K P, and Diaz-Avila M,
{\sl Finite-size scaling of $^4$He at the superfluid transition}, 2008
Rev.\ Mod.\ Phys.\ {\bf 80} 1009

\bibitem{KT}
Kosterlitz J M and Thouless D J,
{\sl Ordering, metastability and phase transitions in two-dimensional systems}
1973  J.\ Phys.\ C {\bf 6} 1181;
Kosterlitz J M, {\sl The critical properties of the two-dimensional XY model},
1974 J.\ Phys.\ C {\bf 7} 1046

\bibitem{Jo77}
Jos\'e J V, Kadanoff L P, Kirkpatrick S and Nelson D R,
{\sl Renormalization, vortices, and symmetry-breaking perturbations in the
two-dimensional planar model}, 1977
Phys.\ Rev.\ B {\bf 16} 1217

\bibitem{AmGoGr80}
Amit D J,  Goldschmidt Y Y and Grinstein G,
{\sl Renormalisation group analysis of the phase transition in the 2D Coulomb
 gas, Sine-Gordon theory and XY model}, 1980
 J.\ Phys.\  A {\bf 13}  585

\bibitem{myKTfilm}
Hasenbusch M,
{\sl Kosterlitz-Thouless transition in thin films:
A Monte Carlo study of three-dimensional lattice models},
2009 J.\ Stat.\ Mech.\ P02005
[arXiv:0811.2178]

\bibitem{myheat}
Hasenbusch M, 
{\sl The specific heat of thin films near the $\lambda$-transition:
A Monte Carlo study of an improved three-dimensional lattice model}, 2009
[arXiv:0904.1535]

\bibitem{BrDaTo00}
J.G. Brankov, D.M. Dantchev, and N.S. Tonchev,
Theory of Critical Phenomena in Finite-Size Systems - Scaling and
Quantum Effects (World Scientific, Singapore, 2000).

\bibitem{GaCh99}
Garcia R and Chan M H W, 
{\sl Critical Fluctuation-Induced Thinning of $^4$He Films near the 
Superfluid Transition}, 1999
Phys.\ Rev.\ Lett. {\bf 83}  1187

\bibitem{GaScGaCh06}
Ganshin A, Scheidemantel S, Garcia R, and Chan M H W, 
{\sl Critical Casimir Force in $^4$He Films: 
Confirmation of Finite-Size Scaling}, 
2006 Phys.\ Rev.\ Lett. {\bf 97} 075301

\bibitem{KrDi92}
Krech M and Dietrich S, {\sl Free energy and specific heat of critical
films and surfaces}, 1992 Phys.\ Rev.\ A {\bf 46} 1886

\bibitem{KrDi92b}
Krech M and Dietrich S, {\sl Specific heat of critical
films, the Casimir force  and wetting films near end points}, 1992
Phys.\ Rev.\ A {\bf 46} 1922

\bibitem{Kardar}
Zandi R, Shackell A, Rudnick J, Kardar M and Chayes L,
{\sl Thinning of superfluid films below the critical point}, 2007
Phys.\ Rev.\ E {\bf  76} (2007)  030601 [cond-mat/0703262]

\bibitem{VaGaMaDi07}
Vasilyev O, Gambassi A, Maciolek A, and Dietrich S,
{\sl Monte Carlo simulation results for critical Casimir forces}, 2007
Europhys.\ Lett.\ {\bf 80} 60009 [arXiv:0708.2902]

\bibitem{Hu07}
Hucht A,
{\sl Thermodynamic Casimir Effect in $^4$He Films near $T_c$:
Monte Carlo Results}, 2007
Phys.\ Rev.\ Lett.\ {\bf 99} 185301 [arXiv:0706.3458]

\bibitem{VaGaMaDi08}
Vasilyev O, Gambassi A, Maciolek A, and Dietrich S,
{\sl Universal scaling functions of critical Casimir forces obtained 
     by Monte Carlo simulations}, 2008
[arXiv:0812.0750]

\bibitem{recentXY}
Campostrini M, Hasenbusch M, Pelissetto A,
and Vicari E,
{\sl
Theoretical estimates of the critical exponents of the superfluid
transition in He4 by lattice methods}, 2006 Phys.\ Rev.\ B {\bf 74} 144506
[cond-mat/0605083]

\bibitem{DiDiEi83}
Diehl H W, Dietrich S, and Eisenriegler E,
{\sl Universality, irrelevant surface operators, and corrections to scaling
in systems with free surfaces and defect planes}, 1983
Phys.\ Rev.\ B {\bf 27} 2937

\bibitem{CaFi76}
Capehart T W and  Fisher M E,
{\sl Susceptibility scaling functions for ferromagnetic Ising films}, 1976
Phys.\ Rev.\ B {\bf 13}  5021

\bibitem{HaTo99}
Hasenbusch M and T\"or\"ok T,
{\sl High precision Monte Carlo study of the 3D XY-universality class},
1999 J.\ Phys.\ A {\bf 32} 6361 [cond-mat/9904408]

\bibitem{ourXY}
Campostrini M, Hasenbusch M, Pelissetto A, Rossi P,
and Vicari E,
{\sl Critical behavior of the three-dimensional XY universality class}, 2001
Phys.\ Rev.\ B {\bf 63} 214503 [cond-mat/0010360]
 
\bibitem{myAPAM}
Hasenbusch M, {\sl
The three-dimensional XY universality class:
A high precision Monte Carlo estimate of the universal amplitude ratio
$A_+/A_-$}, 2006
J.\ Stat.\ Mech.\  P08019 [cond-mat/0607189]

\bibitem{myamplitude}
Hasenbusch M,
{\sl A Monte Carlo study of the three-dimensional XY universality class: Universal
amplitude ratios},  J.\ Stat.\ Mech.\ (2008) P12006 [arXiv:0810.2716]

\bibitem{lipa2003}
Lipa J A, Nissen J A, Stricker D A, Swanson D R and Chui T C P,
{ \sl
Specific heat of liquid helium in zero gravity very near the $\lambda$-point},
2003
Phys.\ Rev.\ B {\bf 68} 174518 [arXiv:cond-mat/0310163]
 
\bibitem{wolff}
Wolff U,
{\sl Collective Monte Carlo Updating for Spin Systems}, 1989
Phys.\ Rev.\ Lett.\ {\bf 62}  361

\bibitem{HaPiVi99}
Hasenbusch M, Pinn K and Vinti S, 
{\sl Critical Exponents of the 3D Ising Universality Class From Finite
Size Scaling With Standard and Improved Actions}, 1999
Phys.\ Rev.\ B {\bf 59} 11471
[hep-lat/9806012]

\bibitem{twister}
Saito M, {\sl An Application of Finite Field:
Design and Implementation of 128-bit Instruction-Based Fast
Pseudorandom Number Generator}, PhD thesis, Dept. of Math.,
Graduate School of Science, Hiroshima University, Advisor: M. Matsumoto;
The numerical program and a detailed description can be found at
``http://www.math.sci.hiroshima-u.ac.jp/{\~{}}m-mat/MT/SFMT/index.html"


\bibitem{RG}
Newman K E and Riedel E K,
{\sl
Critical exponents by the scaling-field method:
The isotropic N-vector model in three dimensions},
1984 Phys.\ Rev.\ B {\bf 30}  6615

\bibitem{BiBhSaBhHu08}
Biswas S,  Bhattacharjee J K, Samanta H S,  Bhattacharyya S,
and  Hu B,
{\sl Theory of the critical Casimir force for He-4 film}
[arXiv:0808.0390] 

\bibitem{webpage}
http://users.wpi.edu/$\sim$garcia/casimirdata/

\bibitem{LiKa91}
Li H and Kardar M, 
 {\sl Fluctuation-Induced Forces between Rough Surfaces}, 1991
Phys.\ Rev.\ Lett. {\bf 67}  3275

\bibitem{Kardar0}
Zandi R,  Rudnick J and Kardar M, 
{\sl Casimir Forces, Surface Fluctuations, and Thinning of Superfluid Film},
2004
Phys.\ Rev.\ Lett. {\bf 93}  155302


\end{thebibliography}
\end{document}